# Saturn's Icy Moon Rhea: a Prediction for its Bulk Chemical Composition and Physical Structure at the Time of the Cassini Spacecraft First Flyby

A. J. R. Prentice[A]

[A]School of Mathematical Sciences,

Monash University, Victoria 3800, Australia

[A]E-mail: andrew.prentice@sci.monash.edu.au

**Abstract**: I report a model for the formation of Saturn's family of mid-sized icy moons to coincide with the first flypast of Rhea by the Cassini Orbiter spacecraft on 26 November 2005. It is proposed that these moons had condensed from a concentric family of orbiting gas rings that were cast off some $4.6 \times 10^9$ yr ago by the proto-Saturnian cloud. The process of condensation and accretion of solids to form Saturn's central planetary core provides the opportunity for the residual gas of the ring to become depleted in rock and $H_2O$ ice relative to the usual solar abundances of these materials. $N_2$, which does not condense, retains its solar abundance relative to $H_2$. If the depletion factor of solids relative to gas is $\zeta_{dep} = 0.25$, as suggested by the low mass of Rhea relative to solar abundance expectations, the mass percent ratio of $NH_3$ to $H_2O$ in the dense proto-Saturnian cloud is $36:64$. Numerical and structural models for Rhea are constructed on the basis of a computed bulk chemical composition of hydrated rock (mass fraction 0.385), $H_2O$ ice (0.395), and $NH_3$ ice (0.220). It is difficult to construct a chemically differentiated model of Rhea whose mean density matches the observed value $\rho_{Rhea} = 1.23 \pm 0.02$ g cm$^{-3}$ for reasonable bounds of the controlling parameters. Chemically homogeneous models can, however, be constrained to match the observed Rhea density provided that the mass fraction of $NH_3$ is permitted to exceed the 'cosmogonic' value of 0.22 by a factor $\zeta_{NH_3} = 1.20 - 1.35$. A large proportion of $NH_3$ in the ice mass inhibits the formation of the dense crystalline phase II of $H_2O$ ice at high pressure. This may explain the lack of compressional features on the surface of the satellite that are expected as a result of ice II formation in the cooling core. The favoured model of Rhea is chemically uniform and has mass proportions: rock (0.369), $H_2O$ ice (0.378) and $NH_3$ ice (0.253). The enhancement factor of $NH_3$ lies within the measured uncertainties of the solar abundance of N. The satellite is very cold and nearly isodense. The predicted axial moment-of-inertia coefficient is $C/MR^2 = 0.399 \pm 0.004$.

# 1  Introduction

*1.1  The Cassini Spacecraft Rhea-1 Flypast and Previous Knowledge of Rhea*

On 26 November 2005, the NASA/ESA Cassini Orbiter spacecraft is to pass within 500 km of the surface of Saturn's second largest moon Rhea. Titan is the largest moon. The Rhea-1 flypast will conclude an exciting series of first-time close encounters with Phoebe, Titan, Enceladus, Hyperion and Dione. The two remaining close encounters with the mid-sized moons do not occur until late in the Cassini mission. These are with Iapetus in September 2007 and Enceladus in March 2008. Each of the Cassini close flypasts are extremely important as they provide a wealth of new data on the physical structure and surficial features of the visited moon. Such data provides a valuable opportunity to test ideas for how Saturn and its family of moons may have formed.

It is the primary purpose of this paper to present a model for the bulk chemical composition and internal structure of Rhea, based on the author's modern Laplacian of solar system origin (Prentice 1978, 1984, 2001). Much of what is presently known about Rhea was obtained 25 years ago from the Voyager 1 & 2 spacecraft encounters (Smith et al. 1981; Tyler et al. 1981). The surface of Rhea is very heavily cratered and appears ancient. That is, there appears to be little evidence of any recent resurfacing events, such is the case for Enceladus. Parts of the surface do, however, display wispy streaks. These are believed to be associated with faults and fractures caused by early episodes of tectonic activity. So far only $H_2O$ ice has been detected on the surface (Clark et al 1984). The absence of $NH_3$ is not surprising owing both to natural sublimation, especially near the equator, and destruction by solar photolysis (Lebofsky 1975; Consolmagno & Lewis 1978). Since the arrival of the Cassini spacecraft at Saturn in July 2004, estimates of the mean radius $R_{Rhea}$ and mass $M_{Rhea}$ have been greatly improved. They are $R_{Rhea} = 764 \pm 4$ km and $M_{Rhea} = (2.306 \pm 0.009) \times 10^{24}$ g. These yield mean density $\rho_{Rhea} = (2.306 \pm 0.009) \times 10^{24}$ g cm$^{-3}$. It is unknown if the interior of Rhea is a homogeneous mixture of rock and ices or if the satellite has differentiated into a rocky core and icy mantle. The answer to that question will emerge from analysis of the gravity data to be acquired during the Cassini flypast of 26 November 2005.

*1.2    Previous Modelling of Rhea*

Previous attempts to model the physical structure of Rhea have been carried out by Consolmagno & Lewis (1978), Lupo & Lewis (1979), Ellsworth & Schubert (1983) and Consolmagno (1985). All these authors have assumed a bulk chemical composition of rock and water ice in the mass percent proportions $60:40$. These proportions follow from the solar abundance compilation of Ross & Aller (1976). Of course, major revisions in the estimates of the solar abundances of the elements have occurred since then (Lodders 2003; Asplund et al. 2005). Today the proportions of rock to ice are close to $50:50$ (see Table 1). This results in an condensate mean density of 1.5 g cm$^{-3}$ which greatly exceeds $\rho_{Rhea}$. Even putting this difficulty to one side, no single consensus for the state of Rhea's interior exists. Consolmagno's work suggests that the melting of $H_2O$ and $NH_3$ ices would occur throughout the inner 40% of the mass of a Rhea-sized moon, so causing differentiation of the rock and ice. Since $NH_3$ ice forms a eutectic melt with $H_2O$ ice at 175 K (Hogenboom et al 1997), the migration of this fluid to the surface would cause partial flooding that might help explain differences in the observed types of cratered terrain.

Ellsworth & Schubert (1983) argue that solid-state convection of the ice would efficiently transfer radiogenic heat to the surface, thus preventing the melting of $H_2O$ ice. Furthermore, even if $NH_3$ ice were present in the compositional mixture as $NH_3$ monohydrate (Lewis 1972), this is only a minor chemical constituent based on solar abundance considerations. Its melting is thus unlikely to substantially modify the viscosity of the mixture or trigger widespread differentiation. That is, the satellite should remain chemically homogeneous. But one main difficulty with an undifferentiated Rhea-sized moon is that the formation of the dense phase II of $H_2O$ ice in the centre, when the satellite cools with the subsidence of radiogenic heating, should have left telltale compressional features at the surface. Such features are not observed. The origin and internal structure of Rhea thus remains a mystery.

## 2  The Modern Laplacian Theory of Solar System Origin

It is proposed that the six mid-sized moons of Saturn, namely Mimas, Enceladus, Tethys, Dione, Rhea and Iapetus condensed from a concentric family of orbiting gas rings that were shed by the proto-Saturnian (hereafter p-Sat) cloud. Gas ring shedding is a central feature of the modern Laplacian theory of solar system origin (Prentice 1978a, 1978b, 1984, 2001). It is the means by which the proto-solar cloud (hereafter PSC) and proto-planetary clouds of Jupiter, Saturn, Uranus and Neptune dispose of excess spin angular momentum during gravitational contraction. The mean orbital radii $R_n$ and masses $m_n$ of the sequence of gas rings ($n = 0, 1, 2, 3...$) satisfy the equations

$$\frac{R_{n-1}}{R_n} \approx \left[1 + \frac{m_n}{M_n f_n}\right]^2 \tag{1}$$

Here $M_n$ and $f_n$ denote the residual mass and moment of inertia factor of the cloud after shedding the $n^{\text{th}}$ gas ring. If the cloud contracts uniformly, so that both $m_n/M_n$ and $f_n$ stay constant, then $R_{n-1}/R_n$ is constant also. That is the orbital radii $R_n$ ($n = 0, 1, 2, 3...$) form a geometric sequence.

We apply equation (1) to the p-Sat cloud and choose $f = 0.01$, $M_n \approx M_{\text{Sat}} = 5.685 \times 10^{29}$ g and $\langle R_{n-1}/R_n \rangle = 1.30$, which is the observed mean orbital distance ratio from Mimas through to Rhea. This yields $m_n = 8.0 \times 10^{26}$ g. We can compare this value with the expected mass of Rhea based on the total mass of condensate (rock, $H_2O$ ice and $NH_3$ ice) in the p-Sat gas ring. Table 1 gives the mass fractions of the broad chemical constituents that can be formed from a gas of solar composition. It is computed on the basis of the proto-solar elemental abundance compilation of Lodders (2003). In constructing this table it is assumed that all rock-like elements have been sequestered into oxides (MgO, $SiO_2$, $Al_2O_3$, $Fe_3O_4$, CaO, $TiO_2$, $Na_2O$, $K_2O$, $Cr_2O_3$,..) sulphides (FeS, NiS, MnS, ZnS) and halides (NaCl, $CaF_2$). Sulphide and halide formation precede that of the oxides. All residual oxygen is used to form $H_2O$. Also all N in the p-Sat gas ring is present as $NH_3$ (Prinn & Fegley 1981). Assuming then that Rhea consists of rock, $H_2O$ ice and $NH_3$ ice in solar proportions, the total mass of condensate is $9.4 \times 10^{24}$ g. This is $\sim 4 M_{\text{Rhea}}$. We propose that the shortfall in the Rhea mass is a natural consequence of the formation of Saturn's central rock/ice core of mass $M_{core} \approx 10-20 M_\oplus$ through the accumulation of the solids which condensed from the proto-solar gas ring at Saturn's orbit. That is, the accretion of Saturn's core may have exhausted most of the condensed chemical species in that gas ring. It is the residual material of that ring (mostly $H_2$ and He) that is later captured by the planetary core to form the p-Sat cloud.

Let $\zeta_{dep}$ denote the depletion factor in the abundance of those elements in the p-Sat cloud that first fully condense as solids in the PSC. This includes all the rock-like elements and that component of total O that does not remain in the gas as $H_2O$ vapour, CO or $CO_2$ (see below and Prentice 1996a, 2001). The mass fraction of element $i$ in the p-Sat envelope is then given by

$$X_{i,Sat} = \zeta_{dep} X_{i,Sun} \qquad (2)$$

We propose that for the p-Sat cloud, $\zeta_{dep} = 0.25$. Table 2 gives the mass fractions of the broad chemical constituents of the p-Sat cloud that now follow from Table 1. In constructing this table, we observe that any element that is not condensed in the proto-solar gas ring retains its solar proportion. This includes the noble gases and N, which exists solely as $N_2$ in the PSC. Also C is distributed between $CH_4$ (number fraction: 0.830), C(s) (0.143), $CO_2$ (0.013) and CO (0.014). Here C(s) means graphite. These numbers emerge from the numerical computation of the gravitational contraction of the PSC that is considered in more detail in Section 3.

## 3    Models for the Proto-solar and Proto-Saturnian Clouds

*3.1    Supersonic Turbulent Stress and the Structure of the Proto-solar cloud*

In order to determine the bulk chemical composition of the native icy moons of Saturn, it is first necessary to construct a numerical model for the PSC. This calculation will yield the thermochemical and compositional state of the gas that makes up the p-Sat cloud. Now in order for the contracting PSC to dispose of its excess spin angular momentum in gas rings, it is necessary that the interior of the cloud be pervaded by a large radial turbulent stress $p_{turb}$ arising from strongly supersonic, thermal convective motions (Prentice 1973; Prentice & Dyt 2003). This stress is given by $p_{turb} = \beta(r)\rho(r)GM(r)/r$. Here $\rho = \rho(r)$ is the local gas density, $M(r)$ is the mass interior to radius $r$ and $\beta = \beta(r)$ is the turbulence parameter. The total pressure at each point is $p_{turb} + p_{gas}$, where $p_{gas} = \rho \Re T/\mu$ is the gas pressure, $T$ is the temperature and $\mu$ is the mean molecular weight.

Each non-rotating model of given surface radius $R_s$ and total mass $M_s = M(R_s)$ has an adiabatic core of radius $r_0$ inside which $\beta = \beta_0$, a constant. The surface of the cloud is defined where the dimensionless temperature function $\theta \equiv \mu_c T(r)/\mu T_c = \theta_s$, a constant, and $c$ refers to the centre. The core itself consists of an inner zone of radius $r_1$ in which all H is taken to be $H_1$ (or $H^+$) and an outer zone in which it is all $H_2$. Lastly, the core is surrounded by a superadiabatic envelope of polytropic index $n_t = -1$ in which $\beta$ falls to 0 as $\theta \to \theta_s$ according as

$$\beta = \frac{\theta - \theta_s}{\theta_0 - \theta_s}, \quad \theta_0 = \frac{\mu_c T(r_0)}{\mu_0 T_c}. \tag{3}$$

Rotation is included using the atmospheric approximation (Prentice 1978a, 1978b). If the controlling parameters $\beta_0, \theta_0, \theta_s$ and $n_t$ stay constant during the contraction, the PSC sheds gas rings whose mean orbital radii $R_n$ ($n = 0, 1, 2, 3,...$) form a nearly geometric sequence. The initial cloud mass $M_t$ is chosen so that the final cloud mass is $M_{Sun} = 1.98892 \times 10^{33}$ g. Setting $\theta_s = 0.002272$, $F_s = \theta_0/\theta_s = 8.534$, and $\beta_0 = 0.1206$ ensures (i) that the mean orbital spacings of the gas rings from Jupiter to Mercury matches the observed planetary spacings, (ii) that the condensate bulk density $\rho_{cond}$ at Mercury's orbit results in a planet of Mercury's physical radius whose mean density equals the observed value, namely $5.43 \pm 0.01$ g/cm$^3$ (Anderson et al. 1987), and (iii) that only a fraction $\phi_{H_2O} = 0.700$ of the total $H_2O$ at Jupiter's orbit condenses out. This ensures a composition for the proto-Jovian cloud that can account for the observed $55:45$ rock-to-ice mass percent ratios in Ganymede and Callisto (Prentice 2001).

At Saturn's orbit, where the PSC gas ring temperature is $T_n = 94$ K, and the mean orbit pressure is $p_n = 4.9 \times 10^{-7}$ bar, the condensate consists of rock (mass fraction: 0.4923), water ice (0.4739) and graphite (0.0338). The rock is almost anhydrous and has density 3.668 g cm$^{-3}$. The total fraction of the water vapour that is condensed is $\phi_{H_2O} = 0.974$. The condensate mean density at the present-day black body temperature at Saturn ($T_{Sat} = 76$ K) is $\rho_{cond} = 1.5225$ g cm$^{-3}$.

*3.2    Gravitational Contraction of the Proto-Saturnian cloud.*

In February 2005, the Cassini Orbiter discovered that Enceladus has a mean density $\rho_{Enc} = 1.606 \pm 0.012$ g cm$^{-3}$ (Rappaport et al. 2005; Anderson et al. 2005). This ~60% higher than a value of 1.00 g cm$^{-3}$ that had been predicted on the basis of a formation model for the Saturn system that had been put forward prior to the Cassini flypast (Prentice 2005a). The primary aim of this pre-Cassini model was to explain the low mean density of Tethys, namely $\rho_{Te} = 0.99 \pm 0.01$ g cm$^{-3}$, relative to its more distant neighbour Dione.

According to Prentice (2005a), when the equatorial radius $R_e$ of the p-Sat cloud shrinks inside the orbit of Dione, the central planetary core of mass $M_{core} \sim 15 M_\oplus$ releases a substantial quantity of its only volatile constituent, namely H$_2$O, into the turbulent p-Sat envelope. Initially the water content of the envelope relative to the solar abundance expectation is $W_{H_2O} = \zeta_{dep} = 0.25$. If some $15-20\%$ of the H$_2$O content of the planetary core is released, the value of $W_{H_2O}$ rises from 0.25 to 3.0. This means that the condensate at the orbit of Tethys now consists mostly of water ice (see Tables 3 & 4). It has a mean density that matches the observed value of Tethys.

*3.3    A Refinement to the p-Sat Cloud Model following the Cassini Close Flypast of Enceladus*

To explain the higher than expected density of Enceladus found by Cassini, it is proposed that the two innermost moons Mimas and Enceladus (and to a lesser extent Tethys) may have initially condensed closer to Saturn than where they are today. These moons then experienced substantial outward radial migration, most possibly as a result of strong tidal action exerted by the young rapidly rotating planet. If Enceladus were to have condensed at orbital distance of $\sim 3.25 R_{Sat}$, rather than at its present distance of $3.951 R_{Sat}$, where $R_{Sat} = 60268$ km, then the temperature $T_n$ of the gas ring (shown in Table 3 and Figure 1) rises from 260 K to 300 K. This value is just below the local condensation temperature of H$_2$O on the mean orbit of the gas ring, namely $T_{H_2O} = 307$ K. This means that most of the water content of the gas ring remains in the vapour phase. As a result, the proportion of rock to H$_2$O in the condensate is greatly enhanced. Because the gas pressure $p_n = 24.5$ bar on the mean orbit is so large, all of the condensing H$_2$O is in the *liquid* state. That is, when Enceladus formed it was a globe of pure water surrounding a rocky core (Prentice 2005b).

Lastly, if we assume that the orbital radii $R_{Di}$ and $R_{Rh}$ of Dione and Rhea were uninfluenced by tidal action, then the present ratio $R_{Rh}/R_{Di} = 1.3967$ can be used to estimate the initial orbital radii $R_{n,0}$ of the family of gas rings shed by the p-Sat cloud during its assumed near-homologous contraction from $30R_{Sat}$ to $3.0R_{Sat}$. These values are shown in Table 3. On this basis, the initial orbital distance of Enceladus is strictly $3.21R_{Sat}$. This difference, however, is not important at this stage. Likewise for Hyperion ($n = 0$), a strict adherence to the geometric spacing law yields $R_{0,0} = 23.8R_{Sat}$. This barely differs from the observed orbital distance $R_0 = 24.29R_{Sat}$, which is adopted in Table 3. Iapetus is assumed to have formed at the $R_{n,0} = 12.2R_{Sat}$ position prior to being displaced to its present orbit as a consequence of Titan's dynamical capture from a solar orbit (Prentice 1984, 2004a, 2004b). Likewise, the moon that once existed at ~$17.0R_{Sat}$ is assumed to have been destroyed collisionally by Titan. Much of the material of that former moon lies buried in Titan's upper mantle and may be the source of Titan's N$_2$–CH$_4$ atmosphere.

*3.4    Specification of the New p-Sat Cloud Model Parameters*

Having described what is expected to be achieved by the new p-Sat cloud model, I now specify the values of the parameters $\beta_0$, $\theta_0$ and $F_s \equiv \theta_s/\theta_0$. First, the value $\theta_{s,Sat} = 0.004822$ ensures that the ratio of the orbital radii of the gas rings shed at the orbital distances of Rhea and Dione matches the observed ratio 1.3967. It should be mentioned here that the mean orbits of all of the native satellites of Saturn underwent a natural radial expansion in the beginning due to the p-Sat cloud losing mass during its gravitational contraction. This process of secular expansion is properly accounted for in the model.

Next, the parameter $\beta_{0,\text{Sat}}$ is standardised against the PSC value $\beta_{0,\text{Sun}}$, assuming a linear dependence on the difference $F_s - 1$. That is, we assume

$$\beta_{0,\text{Sat}} = \beta_{0,\text{Sun}} \cdot \left[ \frac{F_{s,\text{Sat}} - 1}{F_{s,\text{Sun}} - 1} \right] \tag{4}$$

This assumption has a good physical basis since the strength of turbulent stress depends on the temperature contrast between the value at the base of the superadiabatic convecting layer and that at the surface (Prentice & Dyt 2003). If $F_{s,\text{Sat}} - 1 = 0$, all convection ceases. This leaves $F_s = F_{s,\text{Sat}}$ as the residual free parameter of the model. Now $F_s$ controls the absolute scaling of the gas ring temperature distribution $T_n$. During homologous contraction, we have

$$T_n \approx A(F_s)/R_n \tag{5}$$

where the constant $A(F_s)$ depends on $F_s$. The mean density $\rho_{\text{cond}}$ of the condensate at the orbit of Enceladus is extremely sensitive to the choice of $F_s$ since the temperature $T_n$ is so close to the H$_2$O condensation line. Choosing $F_s = 5.6$ we find $\rho_{\text{cond,Enc}} = 1.43 \text{ g cm}^{-3}$, while for $F_s = 5.4$ $\rho_{\text{cond,Enc}} = 1.77 \text{ g cm}^{-3}$, allowing for freezing of the H$_2$O. For $F_s = 5.6$, the density of a frozen Enceladus nearly matches the observed value of 1.606 g cm$^{-3}$. The mean density of the condensate at the orbit of Dione is unchanged as $F_s$ varies from 5.4 to 5.6. The mean density at Rhea's orbit also barely changes. It decreases from 1.250 to 1.247 as $F_s$ increases from 5.4 to 5.6. We therefore choose $F_{s,\text{Sat}} = 5.5$.

*3.5 Specification of the State of Saturn's Central Core and the Other Parameters of the p-Sat Cloud Model*

The remaining aspect of the p-Sat cloud model to specify is the physical state of the central rock/ice core of Saturn. The mass of this core is taken to be $M_{\text{core}} = 10 M_\oplus$. For cloud equatorial radii $R_e \leq 3R_{Sat}$, the density of the core is chosen to be $\rho_{core} = 3.25$ g cm$^{-3}$. It is much beyond the scope of this paper to model the equation of state or evolutionary development of the planetary core. Presumably it underwent considerable compression as the massive p-Sat envelope of mass $\sim 85 M_\oplus$ gathered around it and underwent its own gravitational contraction to a progressively denser state. The initial equatorial radius of the p-Sat cloud is taken to be $R_{e,0} = 30 R_{Sat}$. Prentice (2005a) has proposed that the large water ice content of Tethys is due to the eviction of H$_2$O from the planetary core once the cloud radius $R_e$ has shrunk below $5 R_{Sat}$. In the case of a core of mass $10 M_\oplus$, it is necessary that 25% of the water content of the core be evicted in order to explain the Tethys mean density.

Let $\rho_{core} = \rho_{core}(R_e)$ denote the core density corresponding to cloud equatorial radius $R_e$. A satisfactory model for the core-envelope system is achieved if it is assumed that the initial mass density of the core is $\rho_{core}(30 R_{Sat}) = 0.01$ g cm$^{-3}$. $\rho_{core}$ then increases as a simple power of $R_e$ to the value 1.0 g cm$^{-3}$ at $R_e = 5 R_{Sat}$ and then on up to 3.25 g cm$^{-3}$ at $R_e = 3 R_{Sat}$. It is assumed that the homologous contraction of the p-Sat cloud, dictated by the constancy of the parameters $\beta_0$, $\theta_0$ and $F_s \equiv \theta_s/\theta_0$, terminates at radius $R_{e,*} = 3.0 R_{Sat}$. Below that radius the surface temperature $T_e$ of the cloud is assumed to pass smoothly to a final value of 325 K at radius $R_{e,f} = 1.5 R_{Sat}$. The turbulence parameter $\beta_0$ and polytropic index $n_t$ of the outer superadiabatic layer of the p-Sat cloud also decline in a controlled manner as the final cloud radius is approached. When $\beta \to 0$, $n_t$ passes linearly with $\beta$ to a final value $n_{ad}$. This is the adiabatic value of the gas polytropic index. The decline in turbulent stress causes the geometric spacing between newly shed gas rings to steadily diminish. The Mimas gas ring is shed at $2.777 R_{Sat}$. At radius $R_e = 2.56 R_{Sat}$, all shedding of discrete gas rings ceases. Thereafter the cloud remains rotationally stable by shedding mass continuously, so forming an inner gas disc. Eventually, even disc shedding terminates when the moment-of-inertia factor of the p-Sat cloud rises sufficiently for the rotational velocity at the equator to drop below the Keplerian value.

# 4 Predicted Bulk Chemical Composition of Rhea and a model for its Thermal Evolution

*4.1    Computed Bulk Chemical Compositions of the Satellites*

Table 3 gives the basic physical and chemical properties of the system of gas rings shed by the p-Sat cloud. As noted above, $R_{n,0}$ denotes the initial mean orbital radii of the system of gas rings $(n = 0, 1, 2, 3, ...)$, measured in units of Saturn's present equatorial radius $R_{Sat}$. This table gives the temperature $T_n$ of each gas ring as well as the condensation temperatures $T_i$ of the principal ice species $\{i\}$. These are computed for the mean orbit gas pressures $p_n$. Figure 1 shows $T_n$ and the quantities $T_{H_2O}$, $T_{NH_3}$, etc., plotted as continuous functions of the present orbital distance $R_n$ of a locally detached gas ring. Allowance has been made in this plot for orbital expansion due to mass loss of the p-Sat cloud. $T_{NH_3}$ is the condensation temperature of pure $NH_3$ ice.

Table 4 gives the mass fractions $X_i$ of the principal chemical constituents for each gas ring. $X_{NH_3}$ refers to the total of both pure $NH_3$ ice and $NH_3$ monohydrate. The latter substance is, however, either absent or totally negligible in all moons except Dione. The 2$^{nd}$ last column shows the mean density of the condensate $\rho_{cond}$. This is calculated for the present day solar blackbody temperature at Saturn's orbit. This is $T_{Sat} = 76\,K$ for an albedo of 0.5. The last column of Table 4 gives the observed mean density of the satellite at distance $R_n$ (Jacobson 2004; Jacobson et al. 2005; Rappaport et al. 2005; Anderson et al. 2005).

*4.2    Discussion of the Ammonia-rich Bulk Chemical Composition of Rhea*

The remainder of this paper will be devoted to constructing a viable model for Rhea. The precise mass fractions of the bulk constituents of the Rhea condensate are hydrated rock (0.3853), water ice (0.3946) and pure $NH_3$ ice (0.2201). The most interesting feature of this mix is the high proportion of $NH_3$ ice relative to $H_2O$ ice. If normal solar abundances were to have been assumed for the gas ring from which Rhea condensed, then Table 1 would yield an $NH_3$ to $H_2O$ mass percent ratio of $15:85$. Instead the ratio is $36:64$. That is, $NH_3$ makes up more than $\frac{1}{3}$ of the ice mass. It cannot be ignored. The physical reason for the high proportion of $NH_3$ resides in the proposed 4-fold depletion of the abundance of rock and $H_2O$ ice relative to solar abundances within the residual gas of the proto-solar gas ring at Saturn's orbit. This process was discussed in Section 2. The depletion comes about through the condensation of these species and their subsequent isolation from the gas in forming Saturn's core. The principal chemical constituents of the rock are $SiO_2$ (mass fraction: 0.2795), $Mg(OH)_2$ (0.2760), FeS–NiS (0.1943), $Fe_3O_4$ (0.1010), Fe–Ni (0.0507), $Ca_2MgSi_2O_7$ (0.0391), $MgAl_2O_4$ (0.0298), NaOH–KOH (0.0114). The remaining constituents have mass fractions each less than 0.01. The rock is strongly hydrated and has mean density at 298.15 K and 1 bar of $3.1253\,\text{g cm}^{-3}$.

*4.3    Computation of the Thermal State of Rhea*

A computational code that was designed to model the thermal evolution of the Galilean moons of Jupiter has been applied to Rhea (Prentice 2001). It is assumed that the only heat source within the satellite is that generated by the radioactive decay of $K^{40}$, $Th^{232}$, $U^{235}$ and $U^{238}$ in the rock. Tidal heating is ignored at this stage even although it is possible that the dynamical capture of Titan from a solar orbit (Prentice 1984, 2004a, 2004b) may have greatly stirred the orbital motion of Rhea in the beginning (R.A. Mardling, private communication). The present eccentricity of Rhea is so small that tidal dissipation is practically zero (cf. Peale 1999).

Figure 2 shows the temperature distribution within Rhea as a function of fractional radius $r/R_{\text{Rhea}}$ at a set of key times during the course of the thermal evolution to solar age (4600 Myr). At time $t = 0$, the satellite is assumed to be chemically uniform throughout its interior and have a uniform temperature equal to the present day mean surface value, viz. $T_{\text{surf}} = 76$ K. The profiles in Figure 2 are the solution of the heat diffusion equation assuming that all heat transfer occurs only via conduction. The thermal conductivity of $NH_3$ ice is modelled empirically from the data of Krupskii et al. (1968). It is a temperature dependent quantity, as is the case for the conductivities of the other solid constituents of the Rhea compositional mix. Provided no melting takes place during the evolution, we see that the present day satellite is very cold throughout its interior. The warmest point is the centre, where $T_c(4600\,\text{Myr}) = 112$ K.

*4.4    Examining the Possibility of Melting of Ice*

Consider now the important question of whether any melting of ice takes place at some point during the thermal evolution to present age. The melting temperatures $T_{m,i}$ of $H_2O$ ice and $NH_3 \cdot 2H_2O$ ice are shown in Figure 2. Both of these quantities are functions of pressure $p$ and hence of the fractional radius $r/R_{\text{Rhea}}$. For ammonia dihydrate $(i = 2)$, Hogenboom et al. (1997) find

$$T_{m,2}(p) = 176.29 + 2.43 \times 10^{-3} p - 7.7583 \times 10^{-5} p^2 \text{ K,}$$

where the unit of $p$ is bar. At the centre of the satellite, where $p_c = 1234$ bar in the thermal code, $T_{m,2} = 178$ K. Not shown in Figure 2 is $T_{m,3}$ for pure $NH_3$ ice. This everywhere about 20 K above that of the dihydrate ice. However, when $H_2O$ and $NH_3$ coexist, melting first occurs at the eutectic point for the mixture of these ices. This $T_{m,2}$.

It is clear from Figure 2 that if the possibility of convective heat transfer through solid state creep of ice is ignored, then melting will take place for all fractional radii $r/R_{\text{Rhea}} \leq 0.65$ K. This first occurs after ~500 Myr. Ellsworth & Schubert (1983) were the first to explicitly model convective heat transfer in icy satellites. According to these authors, the Rayleigh number quickly rises above a critical value ~1100 required for the onset of convection in a Rhea-sized radiogenically heated icy satellite. Even so, convection cannot start until the temperature exceeds a fixed fraction $f_{\text{creep}}$ of the local melting temperature $T_m$. The curve marked $T_{\text{creep}}$ in Figure 2 is the solid-state creeping temperature of $H_2O$ ice for the case $f_{\text{creep}} = 0.7$. Ellsworth & Schubert (1983) adopted $f_{\text{creep}} = 0.6$. Prentice & Freeman (1999) found that a higher value is needed if we are to account for the existence of a sub-surface layer of liquid ammonia dihydrate in Callisto.

In the case of Callisto, NH$_3$ ice makes up only ~9% of the ice mass, so the rheology is controlled by the properties of H$_2$O ice. For model of Rhea proposed here, however, NH$_3$ comprises ~36% of the ice. This closely matches the dihydrate proportion (32%). This means that the solid state convection is now controlled by the rheology of the dihydrate ice. The true creep temperature will thus be closer to ~140 K. We can be fairly certain, therefore, that none of the interior of Rhea will undergo melting at any stage during its thermal evolution to solar age if all heating is derived solely from radiogenic decay. The present temperature distribution within Rhea may thus be reasonably assumed to be very similar to the $t = 4600 \, \text{Myr}$ profile shown in Figure 2.

## 5 Structural Models of Rhea

*5.1 The Controlling equations and the Equations of State*

It remains to construct a self-consistent model for Rhea, based on the computed temperature profile found in Section 4. This task is performed using a computational code that was also developed to numerically model the interior structures of the Galilean satellites (Prentice 1996a, 1996b, 2001).

The distribution of pressure $p = p(r)$ with radius $r$ in a spherically symmetric satellite of surface radius $R_s$ is controlled by the hydrostatic support equation

$$\frac{dp}{dr} = -\frac{\rho G M(r)}{r^2}.$$

Here $\rho = \rho(r)$ is the local mean density and $M(r)$ is the mass interior to radius $r$. For a solid composed of $N$ chemical species of mass fraction $X_i$ and density $\rho_i$, the mean density is given by

$$\frac{1}{\rho} = \sum_{i=1}^{N} \frac{X_i}{\rho_i}.$$

Each of the quantities $\rho_i$ depends on both $p$ and the local temperature $T$. That is, it is necessary to know the equation of state (EOS) of each of the chemical constituents of the solid. In the case of most rocky minerals, an excellent representation of the thermal and compressional behaviour is achieved by fitting a Murnaghan EOS. We have

$$\rho_i(p,T) = \rho_{i,1}(T) \cdot \left[1 + \frac{K'_{1,i}}{K_{1,i}}(p-1)\right]^{\alpha}, \quad \alpha = \frac{1}{K'_{1,i}}.$$

Here $K_{1,i}$ and $K'_{1,i}$ denote the bulk modulus and its first pressure derivative at 1 bar pressure and $\rho_{i,1}(T)$ is the mean density also at 1 bar. The main advantage of this separable representation is that one can utilize the great body of experimental data on the thermal expansion of solids that exists for most minerals at 1 bar pressure.

The EOS for water ice is extremely complex owing to the existence of the many high pressure phases of this ice. Empirical representations for the densities $\rho_{H_2O,j}(p,T)$ of each phase $j$ are assembled in Prentice (2001). For $NH_3$ ice the following Murnaghan-style EOS has been constructed by combining the thermal expansion data of Croft et al. (1988) and compresssional data of Stewart (1960).

$$\rho_{NH_3,1}(T) = 0.8659\, e^{-4.198 \times 10^{-7} T^{2.2207}} \quad \text{g cm}^{-3}$$

$$A_1 = (1.495 + 0.001T) \times 10^{-5}, \quad B_1 = (4.5066 - 7.778 \times 10^{-3} T) \times 10^{-10}$$

$$K'_{NH_3,1} = 2B_1/A_1^2 - 1, \quad K_{NH_3,1} = 1/A_1 \quad \text{bar}$$

*5.2    Results and Discussion*

Five distinct models for Rhea have been constructed according to the choice of the various controlling parameters which are listed in Table 5. These parameters include the global mass fractions $X_i$ of rock, $H_2O$ ice and $NH_3$ ice, and a new quantity $\zeta_{NH_3}$. This defines an enhancement of the mass of $NH_3$ relative to the cosmogonically derived value of 0.2201 given in Table 4. The 5 models are labelled H1, D, H2, H3 and H4, respectively. Models H1, H2, H3 and H4 have chemically homogeneous interiors. Model D is a differentiated structure having a central rocky core and icy mantle. Each satellite model consists of two zones of different temperature $T_1$ and $T_2$. For the homogeneous models, the boundary between these zones is the mass mid-point. The last column gives the transition pressure $p_{I/II}$ that defines the boundary between phase I and II of water ice. All quantities have been computed for the thermal profile at 4600 Myr given in Figure 2.

*5.2.1  A First Homogeneous Satellite Model: H1*

This model has a uniform mixture of rock and ices in the proportions established from solar abundances and the depletion factor $\zeta_{dep} = 0.25$ that was introduced in Section 2. The principal characteristics of the model are listed in Table 6. Within the inner ~17% of the satellite mass, all H₂O ice is in the dense phase II state. This phase is ~27% denser than phase I ice at the same pressure. Overall, the model is too dense compared to the observed density of Rhea, namely, $\rho_{Rhea} = 1.23 \pm 0.02$ g cm$^{-3}$, the error being $2.5\sigma$.

*5.2.2  A Differentiated Satellite Model: D*

This model assumes that all rock has settled to the centre following an unanticipated meltdown of all the ice. If such a meltdown did take place, perhaps as a result of powerful tidal action due to Titan, the liquid mantle would refreeze as a eutectic mixture forming a deep mantle of NH₃·2H₂O ice of mass fraction 0.5811. This mantle is surmounted by a thin crust of pure NH₃ ice of mass fraction 0.00336 and thickness 13.5 km. Remarkably, the mean density of the model coincides with the value $1.33 \pm 0.10$ g cm$^{-3}$ obtained by the Voyager 1 spacecraft (Tyler et al. 1981). The Cassini data suggests that this model is not a viable one.

*5.2.3  A Second Homogeneous Model: H2*

This model attempts to match the observed mean density of Rhea by assuming that the actual mass of NH₃ ice relative to rock and H₂O ice is a factor $\zeta_{NH_3} = 1.35$ times larger than the value suggested earlier. Now the solar abundance of N has an uncertainty of 29% (Lodders 2003). The $\zeta_{NH_3}$ value for model H2 thus lies just outside the error range, but not significantly so. We need to bear in mind that the value of the depletion factor $\zeta_{dep}$ relating to the exhaustion of solids relative to gas in the proto-solar gas ring is also uncertain, probably by ~20%. That is $\zeta_{dep} = 0.25 \pm 0.05$. The H2 model thus cannot be ruled out on abundance grounds.

*5.2.4  Homogeneous Models without Water Ice Phase II: Models H3 and H4*

The only objection to Model H1 was that it is too dense. Now this satellite model contains a large quantity of the dense phase II of $H_2O$ at its centre. As Ellsworth & Schubert (1983) have noted, the formation of this material in the core of the cooling satellite should have caused the mean radius to slump by about 15 km. In our model the estimated slump is ~5 km. Even so, such shrinkage should have left prominent compressional features on the surface, such as networks of parallel ridges and troughs. Such features have so far not been observed, suggesting that the formation of phase II of water ice was somehow suppressed.

The crucial step now is the realisation that because the ice in the present study consists of an intimate mixture of $H_2O$ and $NH_3$ of comparable proportions, the formation of phase II ice may never take place. That is, the massive intrusion of $NH_3$ may prevent the crystallization of the dense phases of water ice. This idea was first mooted by D.J. Stevenson, as a private communication in Consolmagno (1985).

The remaining structural models of Rhea do not possess any ice II, despite the fact that the pressure exceeds the $p_{I/II}$ transition pressure throughout much of the central region of the satellite. Model H3 assumes the standard bulk composition of rock, $H_2O$ ice and $NH_3$ ice, corresponding to the choice $\zeta_{NH_3} = 1.0$. For model H4, $\zeta_{NH_3} = 1.2$. The mean density of the H3 model is much closer to the observed value of Rhea. In fact it lies only just over $1\sigma$ away. The best fitting of all the Rhea models is H4. Here $\zeta_{NH_3}$ is chosen so that the model mean density matches $\rho_{Rhea}$. The required increase in $\zeta_{NH_3}$ from 1.0 to 1.2 is well inside the observational uncertainty of the solar N abundance. It is also compatible with the permitted range of the depletion factor $\zeta_{dep} = 0.25 \pm 0.05$.

Finally, since the compression of rock and ice is everywhere so modest, there is very little contrast between the central and surface densities of the satellite model. The temperature contrast between these regions is also small. It is predicted that the Cassini measurements taken during the Rhea-1 close flypast of 26 November 2005 will find Rhea to be a cold and essentially chemically homogeneous structure. The predicted axial moment-of-inertia coefficient is

$$[C/MR^2]_{\text{Rhea}} = 0.399 \pm 0.004.$$

## 6 Conclusions

Theoretical and numerical models for the origin, bulk chemical composition and internal structure of Rhea have been put forward at the time of the Cassini spacecraft first close flyby. It is proposed that as a result of condensation and depletion of solids within the proto-solar gas ring from which the p-Sat gas cloud was derived, that this cloud was deficient by a factor $\zeta_{\text{dep}} = 0.25$ in rock and $H_2O$ ice relative to solar abundance of these materials. Nitrogen, however, is present in nearly solar proportion to $H_2O$ and, as a result, the mass percent ratio of $NH_3$ to $H_2O$ in the p-Sat cloud is 36:64.

It is proposed that Saturn's mid-sized moons condensed from a concentric family of gas rings that were cast off at the equator of the contracting p-Sat cloud. A process of supersonic turbulent convection that has previously been applied successfully to calculate the properties of the system of gas rings shed by the proto-solar and proto-Jovian clouds is used to determine the properties of the p-Sat gas rings. The temperatures $T_n$ of these rings vary closely with mean orbital distance $r_n$ ($n = 0, 1, 2, 3, ...$) according as $T_n = A/R_n$. If the constant of proportionality is chosen to account for the observed mean density of Enceladus, then Rhea condenses well inside the stability field of $NH_3$ ice. The predicted bulk chemical composition is: rock (mass fraction 0.3853), $H_2O$ ice (0.3946) and $NH_3$ ice (0.2201). We refer to this as the 'cosmogonic' chemical mixture.

A family of numerical models for the thermal evolution and present-day interior structure of Rhea have been constructed. It is not possible to construct a chemically differentiated satellite model whose mean density matches that of Rhea. Homologous models can be constrained to match the observed mean density provided that it is assumed that the mass fraction of $NH_3$ ice exceeds the 'cosmogonic' value by a factor $\zeta_{NH_3} = 1.20 - 1.35$. Next it is proposed that the large quantity of $NH_3$ ice, relative to solar abundance expectation, may have prevented the formation of the dense phase II of $H_2O$ ice at the centre of Rhea. This may explain the observed dearth of compressional features on the surface. Such features would have otherwise formed as the central regions of the moon cooled following the subsidence of radiogenic heating.

The favoured model of Rhea has a chemically uniform interior and is very cold. The central temperature exceeds the surface value by barely than 40 K. The satellite is nearly isodense and the predicted value of the axial moment-of-inertia factor is $[C/MR^2]_{Rhea} = 0.399 \pm 0.004$. $NH_3$ is unstable at Saturn's distance from the Sun except near the polar regions of a satellite. Perhaps the Cassini Orbiter will discover indirect evidence for $NH_3$ through the sublimative escape of this ice from the outer layers, especially near the equatorial zones. Wasting of $NH_3$ would weaken the residual soil, so making the edges of craters soft and prone to landslides. It will be exciting to learn what Cassini discovers.

**Acknowledgements**

The author thanks P. D. Godfrey, R. A. Mardling, and N.J. Rappaport for helpful discussions.

G. W. & C. Null, C. & M. Savalla, and E.M. Standish offered generous hospitality in Pasadena.

L. Mayer, C. Morgan, S. Morton, A. Thorne, and C. Wilson provided valuable technical support.

**References**

Anderson, J. D., Colombo, G., Esposito, P. B., Lau, E. L., & Trager, G. B. 1987, Icarus, 71, 337

Anderson, J. D., Schubert, G., Anatbtawi, A., Asmar, S.W., Iess, L., Rappaport, N.J., Somenzi, L., Tortora, P., & Zingoni, F. 2005, Science, in press

Asplund, M., Grevesse, N., & Sauval, A. J. 2005. in ASP Conf. Ser. 336: Cosmic Abundances as Records of Stellar Evolution and Nucleosynthesis, eds. F. N. Bash and T. G. Barnes (Provo: Brigham Young University), 25

Clark, R. N., Brown, R. H., Owensby, P. D., & Steele, A. 1984, Icarus, 58, 265

Consolmagno, G. J. 1985, Icarus, 64, 401

Consolmagno, G. J., & Lewis, J. S. 1978, Icarus, 34, 280

Croft, S. K., Lunine, J. I., & Kargel, J. 1988, Icarus, 73, 279

Ellsworth, K., & Schubert, G. 1983, Icarus, 54, 490

Hogenboom, D. L., Kargel, J. S., Consolmagno, G. J., Holden, T. C., Lee, L., & Buyyounouski, M. 1997, Icarus, 128, 171

Jacobson, R. A. 2004, Astron J, 128, 492

Jacobson, R. A., Antreasian, P. G., Bordi, J. J., Criddle, K. E., Ionasescu, R., Jones, J. B., Mackenzie, R. A., Meek, M. C., Pelletier, F. J., Roth, D. C., Roundhill, I. M., & Stauch, J. R. 2005, BAAS, 36, 524

Krupskii, I. N., Manzhely, V. G., & Koloskova, L.A., 1968, Phys. Stat. Sol., 27, 263

Lebofsky, L. A. 1975, Icarus, 25, 205

Lewis, J. S. 1972, Icarus, 16, 241

Lodders, K. 2003, ApJ, 591, 1220

Lupo, M. J., & Lewis, J. S. 1979, Icarus, 40, 157

Peale, S. J. 1999, ARA&A, 37, 533

Prentice, A. J. 1973, A&A, 27, 237

Prentice, A. J. R. 1978a, in Origin of the Solar System, ed. S. F. Dermott (New York: John Wiley),

Prentice, A. J. R. 1978b, Moon & Planets, 19, 341

Prentice, A. J. R. 1984, Earth, Moon & Planets, 30, 209

Prentice, A. J. R. 1996a, Earth, Moon & Planets, 73, 237

Prentice, A. J. R. 1996b, Phys. Letts., A213, 253

Prentice, A. J. R. 2001, Earth, Moon & Planets, 87, 11

Prentice, A. J. R. 2004a, BAAS, 36, 780

Prentice, A. J. R. 2004b, BAAS, 36, 1116

Prentice, A. J. R. 2005a, Lunar Planet Sci., 36, 2378

Prentice, A. J. R. 2005b, BAAS, 37, 729

Prentice, A. J., & Freeman, J. C. 1999, Eos Trans AGU, 80, F607

Prentice, A. J. R., & Dyt, C. P. 2003, MNRAS, 341, 644

Prinn, R. G., & Fegley, B. 1981, ApJ, 249, 308

Rappaport, N. J., Iess, L., Tortora, P., Asmar, S. W., Somenzi, L., Anabtawi, A., Barbinis, E., Fleischman, D. U., & Goltz, G. L. 2005, BAAS, 37, 704

Ross, J. E., & Aller, L. H. 1976, Science, 191, 1223

Smith, B. A., Soderblom, L., Beebe, R., Boyce, J., Briggs, G., Bunker, A., Collins, S. A., Hansen, C. J., Johnson, T. V., Mitchell, J. L., Terrile, R. J., Carr, M., Cook, A. F., Cuzzi, J., Pollack, J. B., Danielson, G. E., Ingersoll, A., Davies, M. E., Hunt, G. E., Masursky, H., Shoemaker, E., Morrison, D., Owen, T., Sagan, C., Veverka, J., Strom, R., & Suomi, V. E. 1981, Science, 212, 163

Stewart, J. W. 1960, J.Chem.Phys, 33, 128

Tyler, G. L., Eshleman, V. R., Anderson, J. D., Levy, G. S., Lindal, G. F., Wood, G. E., & Croft, T. A. 1981, Science, 212, 201

**Table 1: Broad Chemical Composition of Proto-Solar Material***

| Category | Species | Mass fraction |
|---|---|---|
| Gases | $H_2$ | 0.710864 |
| | He | 0.272684 |
| | Ne & Ar | 0.001379 |
| Ices | $H_2O$ | 0.005368 |
| | $NH_3$ | 0.000966 |
| | $CH_4$ | 0.003302 |
| Rocks | See text | 0.005437 |
| Total | | 1.000000 |

*Derived from the solar elemental abundance date of Lodders (2003)

**Table 2: Broad Chemical Composition of the Proto-Saturnian Cloud**

| Category | Species | Mass fraction |
|---|---|---|
| Gases | $H_2$ | 0.716891 |
| | He | 0.274958 |
| | Ne & Ar | 0.001391 |
| Ices | $H_2O$ | 0.001460 |
| | $NH_3$ | 0.000974 |
| | $CH_4$ | 0.002955 |
| Rocks | Same as Table 1 | 0.001371 |
| Total | | 1.000000 |

**Table 3: Properties of the Proposed Family of Gas Rings Shed by the Proto-Saturnian Cloud**

| Moon | Index $n$ | $R_n$ ($R_{Sat}$) | $R_{n,0}$ ($R_{Sat}$) | $p_n$ (bar) | $T_n$ (K) | $T_{H2O}$ (K) | $T_{NH3}$ (K) | $T_{CH4 \cdot 5.75H2O}$ (K) |
|---|---|---|---|---|---|---|---|---|
| Mimas | 7 | 3.080 | 2.777 | 33.6 | 317 | 328 | 172 | 145 |
| Enceladus | 6 | 3.951 | 3.250 | 24.5 | 300 | 307 | 170 | 142 |
| Tethys | 5 | 4.890 | 4.484 | 7.87 | 222 | 288 | 161 | 132 |
| Dione | 4 | 6.262 | 6.262 | 2.65 | 170 | 245 | 154 | 123 |
| Rhea | 3 | 8.746 | 8.746 | 0.809 | 129 | 234 | 147 | 115 |
| Iapetus | 2 | 12.20 | 12.20 | 0.241 | 101 | 224 | 140 | 108 |
| ? | 1 | 17.00 | 17.00 | 0.070 | 83 | 215 | 134 | 102 |
| Hyperion | 0 | 24.29 | 24.29 | 0.0176 | 74 | 205 | 128 | 95 |

**Table 4: Condensate Bulk Composition and Mean Density**

| Moon | $R_{n,0}$ ($R_{Sat}$) | $W_{H2O}$ | $X_{rock}$ | $X_{H2O}$ | $X_{NH3}$ | $X_{CH4}$ | $\rho_{cond}$ (g cm$^{-3}$) | $\rho_{obs}$ (g cm$^{-3}$) |
|---|---|---|---|---|---|---|---|---|
| Mimas | 2.777 | 6.5 | 0.277 | 0.723 (L) | 0. | 0. | 1.16 | $1.17 \pm 0.04$ |
| Enceladus | 3.250 | 3.0 | 0.574 | 0.426 (L) | 0. | 0. | 1.57 | $1.60 \pm 0.02$ |
| Tethys | 4.484 | 3.0 | 0.088 | 0.912 | 0. | 0. | 0.99 | $0.96 \pm 0.01$ |
| Dione | 6.262 | 0.25 | 0.494 | 0.506 | *$5 \times 10^{-5}$ | 0. | 1.43 | $1.47 \pm 0.02$ |
| Rhea | 8.746 | 0.25 | 0.385 | 0.395 | 0.220 | 0. | 1.25 | $1.23 \pm 0.01$ |
| Iapetus | 12.20 | 0.25 | 0.351 | 0.360 | 0.233 | 0.056 | 1.14 | $1.11 \pm 0.04$ |
| ? | 17.00 | 0.25 | 0.338 | 0.367 | 0.238 | 0.057 | 1.14 | |
| Hyperion | 24.29 | 0.25 | 0.333 | 0.370 | 0.240 | 0.057 | 1.14 | $0.58 \pm 0.11$ |

(L) means liquid water;  *mass fraction of $NH_3$ tied up as $NH_3 \cdot H_2O$

## Table 5: Controlling Parameters for the Family of Rhea Structural Models

| Model | $X_{rock}$ | $X_{H_2O}$ | $X_{NH_3}$ | $\zeta_{NH_3}$ | $T_1$ (K) | $T_2$ (K) | $p_{I/II}$ (bar) |
|---|---|---|---|---|---|---|---|
| H1 | 0.3853 | 0.3946 | 0.2201 | 1.00 | 96 | 81 | 960 |
| D | 0.3853 | 0.3946 | 0.2201 | 1.00 | 100 | 81 | 878 |
| H2 | 0.3577 | 0.3664 | 0.2759 | 1.35 | 97 | 81 | 976 |
| H3 | 0.3853 | 0.3946 | 0.2201 | 1.00 | 96 | 81 | – |
| H4 | 0.3690 | 0.3780 | 0.2530 | 1.20 | 97 | 81 | – |

## Table 6: Structural Properties of the Rhea Models

| Property | H1 | D | H2 | H3 | H4 |
|---|---|---|---|---|---|
| Central pressure (bar) | 1435 | 3454 | 1304 | 1289 | 1241 |
| Rock core radius (km) | – | 416 | – | – | – |
| Radius of dihydrate boundary (km) | – | 750 | – | – | – |
| Ice I / ice II boundary (km) | 410 | – | 356 | – | – |
| Ice II / Total $H_2O$ ice | 0.171 | – | 0.112 | – | – |
| Central density (g cm$^{-3}$) | 1.425 | 3.172 | 1.362 | 1.264 | 1.240 |
| Surface density (g cm$^{-3}$) | 1.249 | 0.980 | 1.210 | 1.245 | 1.226 |
| Mean density (g cm$^{-3}$) | 1.279 | 1.330 | 1.229 | 1.254 | 1.231 |
| $C/MR^2$ | 0.394 | 0.324 | 0.396 | 0.3995 | 0.3995 |

**Figures and Figure Captions**

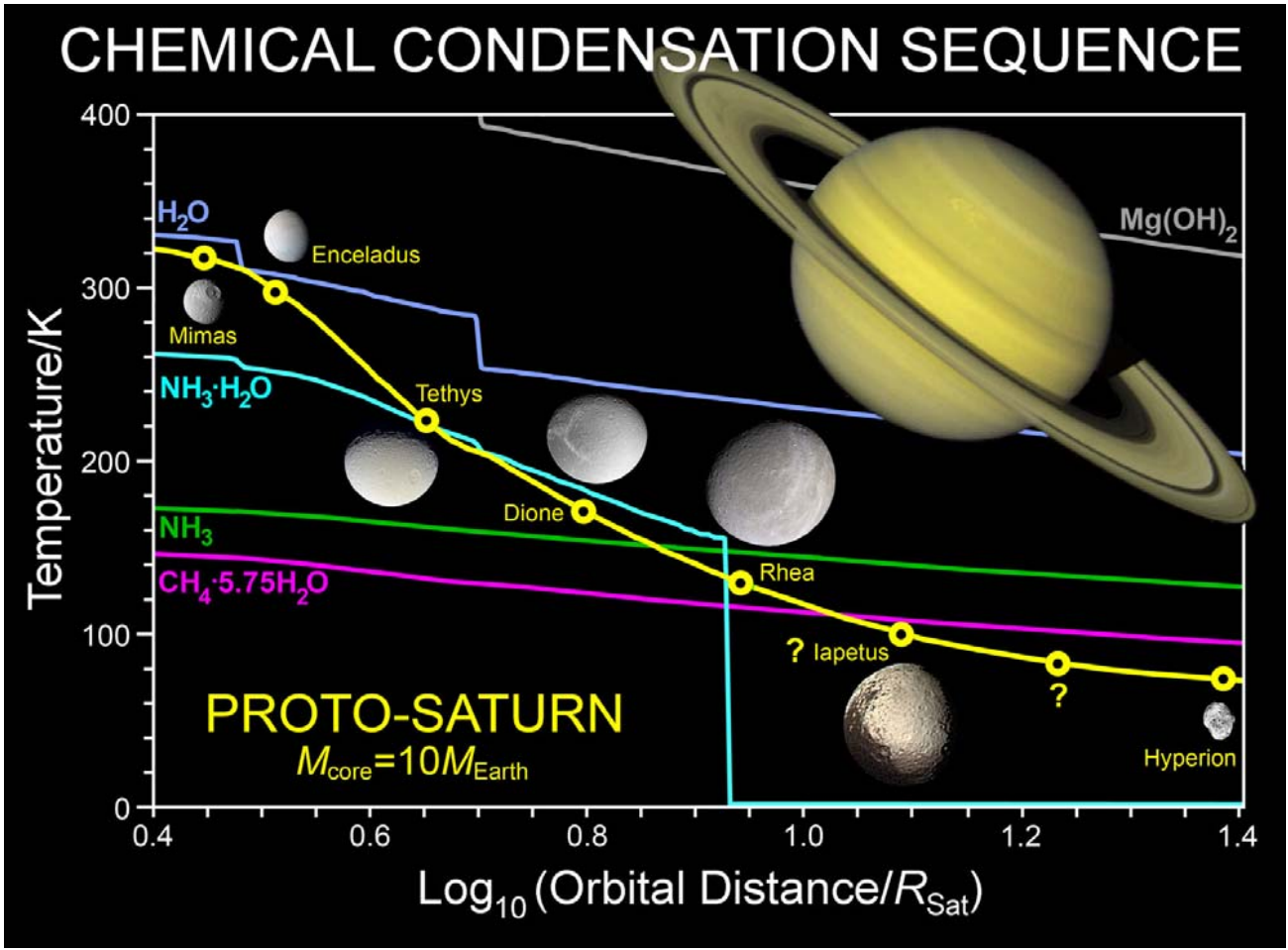

**Figure 1** The heavy yellow curve in this diagram gives the temperature of a gas ring at the moment of detachment from the contracting proto-Saturnian (p-Sat) cloud. It is calculated against the present orbital radius that the ring would have if it existed today, allowing for secular expansion due to mass loss from the p-Sat cloud. The unit of radius is $R_{Sat} = 60268$ km. The temperatures $T_n$ of the gas rings from which Saturn's native moons condensed are also plotted against their present orbital distances $R_n$ $(n = 0,1,2,3,...)$. The open circles show them. The other lines in the Figure give the condensation temperatures $T_i$ of the principal ice species $\{i\}$ as well as that of the copious rocky constituent brucite. These temperatures are calculated for the gas pressure $p_n$ on the mean orbit of the gas ring. The values of $T_i$, $R_n$ and $p_n$ for each of the satellites are shown in Table 3. The abrupt changes in $T_{H_2O}$ at orbital distances $3R_{Sat}$ and $5R_{Sat}$ are due to an enhancement in the water vapour content of the p-Sat envelope. This occurs through a proposed release of $H_2O$ by Saturn's core of mass $M_{core} = 10 M_\oplus$, as discussed in the text.

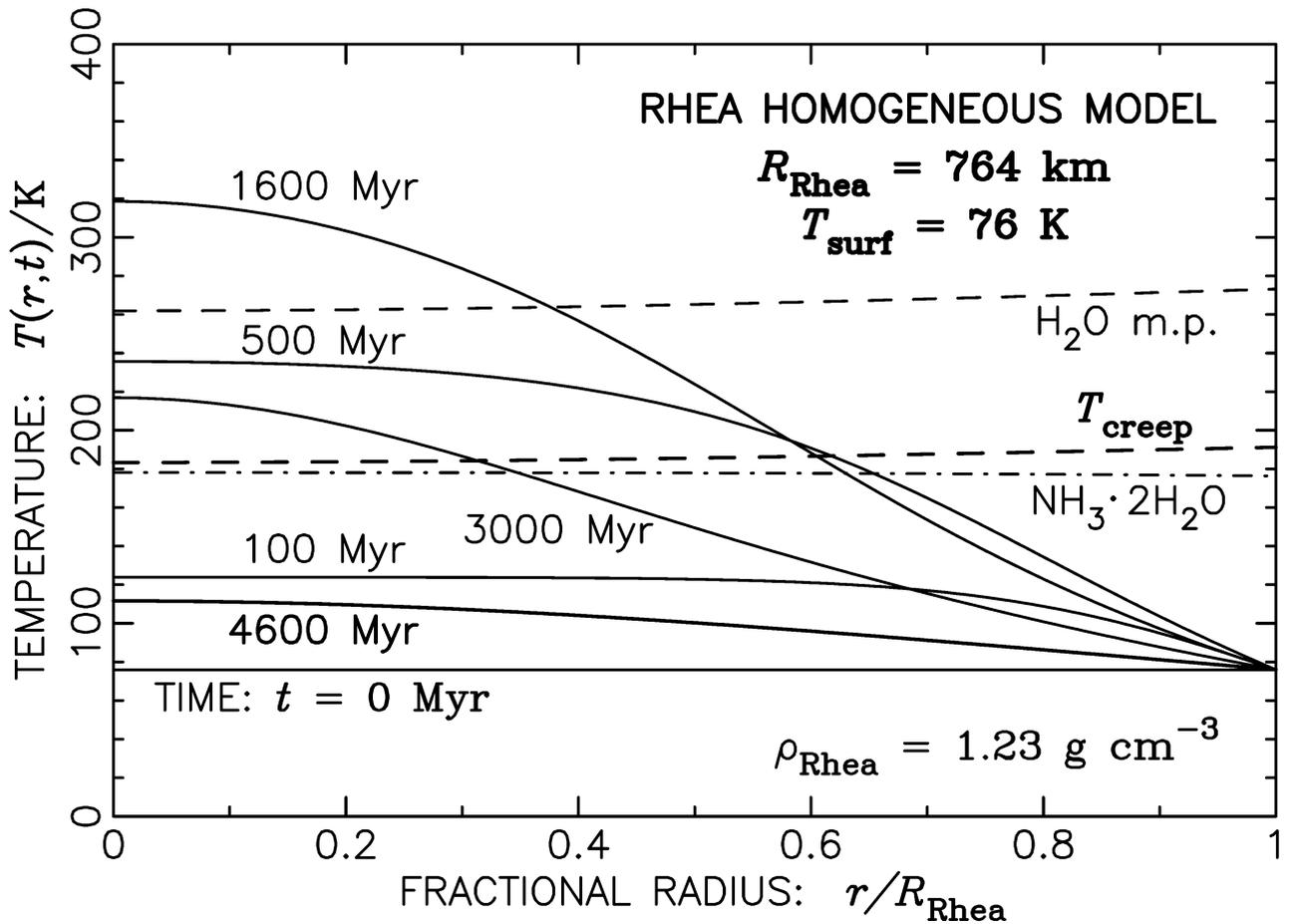

**Figure 2** Internal temperature profiles of a homogeneous Rhea-sized satellite comprised of rock, H$_2$O and NH$_3$ ice in the fractional mass proportions 0.3853, 0.3946 and 0.2201. The profiles of temperature versus fractional radius $r/R_{Rhea}$ are shown at various times $t$ during the course of the satellite's thermal evolution to present age (4600 Myr). All heat is derived from the decay of radioactive nuclides. The melting temperature of H$_2$O ice and that of ammonia dihydrate are shown by the broken lines. They depend on pressure and hence on $r/R_{Rhea}$. The satellite becomes locally unstable towards solid-state convection when the temperature $T(r,t)$ rises through the creep value $T_{creep} = 0.7T_{H_2O}$. The contribution of solid-state convection to the heat transfer rate has not been included here.